\documentclass{aa}
\usepackage{epstopdf}
\usepackage{txfonts}
\bibpunct{(}{)}{;}{a}{}{,} 
\begin{document}

\title{An ALMA Survey for Disks Orbiting Low-Mass Stars in the TW Hya Association}

\author{David R.\ Rodriguez\inst{1}
\and Gerrit van der Plas\inst{1,2}
\and Joel H.\ Kastner\inst{3}
\and Adam C.\ Schneider\inst{4}
\and Jacqueline K.\ Faherty\inst{5,6}
\and Diego Mardones\inst{1}
\and Subhanjoy Mohanty\inst{7}
\and David Principe\inst{2,8}
}

\institute{
Departamento de Astronom\'ia, Universidad de Chile, Casilla 36-D, Santiago, Chile 
\email{drodrigu@das.uchile.cl} 
\and
Millennium Nucleus Protoplanetary Disks, Chile
\and
Center for Imaging Science, School of Physics \& Astronomy, and Laboratory for Multiwavelength Astrophysics, Rochester Institute of Technology, 54 Lomb Memorial Drive, Rochester, NY 14623, USA
\and
Department of Physics and Astronomy, University of Toledo, 2801 W. Bancroft St., Toledo, OH 43606, USA
\and
Department of Terrestrial Magnetism, Carnegie Institution of Washington, 5241 Broad Branch Road NW, Washington, DC 20015, USA
\and 
Hubble Fellow
\and
Imperial College London, 1010 Blackett Lab., Prince Consort Road, London SW7 2AZ, UK
\and
N\'ucleo de Astronom\'ia, Facultad de Ingenier\'ia, Universidad Diego Portales, Av. Ejercito 441, Santiago, Chile
}

\abstract
{
We have carried out an ALMA survey of 15 confirmed or candidate low-mass ($<0.2M_\odot$) members of the TW Hya Association (TWA) with the goal of detecting molecular gas in the form of CO emission, as well as providing constraints on continuum emission due to cold dust.
Our targets have spectral types of M4-L0 and hence represent the extreme low end of the TWA's mass function. 
Our ALMA survey has yielded detections of 1.3mm continuum emission around 4 systems (TWA~30B, 32, 33, \& 34), suggesting the presence of cold dust grains. All continuum sources are unresolved. TWA~34 further shows $^{12}$CO(2--1) emission whose velocity structure is indicative of Keplerian rotation. 
Among the sample of known $\sim$7--10 Myr-old star/disk systems, TWA~34, which lies just $\sim$50 pc from Earth, is the lowest mass star thus far identified as harboring cold molecular gas in an orbiting disk.
}

 
\titlerunning{A Molecular Disk Survey of Low-Mass Stars in the TW Hya Association}
\authorrunning{D.R. Rodriguez et al.}

\maketitle

\section{Introduction}

Beginning with the identification of the TW Hya Association (TWA; \citealt{Kastner:1997}) several stellar associations with ages $\sim$8--200~Myr have been identified in close proximity to the Earth \citep{ZS04,Torres:2008}. 
These young moving groups serve as excellent laboratories to explore the evolution of stellar and planetary properties (e.g., \citealt{Marois:2008, Lagrange:2010, Rodriguez:2010}). 
Of particular interest is the growing number of M-dwarfs identified as members of these groups and how their properties, such as disk lifetimes, compare with those of higher mass members. 

Surveys of star forming regions have shown that protoplanetary disk lifetimes are on the order of $\sim$2--3~Myr \citep{Williams:2011}.
To date, we know only of a handful of pre-main sequence stars within 100~pc of Earth that, despite having ages of $\sim$5--20~Myr, still host disks that appear primordial in nature.
These include TW~Hya ($\sim$8~Myr, $\sim$50~pc; \citealt{Kastner:1997}), V4046~Sgr ($\sim$20~Myr, $\sim$70~pc; \citealt{Kastner:2008b}), MP~Mus ($\sim$5--7~Myr, $\sim$100~pc; \citealt{Kastner:2010}), and T~Cha ($\sim$5--7~Myr, $\sim$110~pc; \citealt{Sacco:2014}). 
All of these are roughly solar-mass (K type) stars; whether or not lower-mass (M type) stars can retain primordial disks to such relatively advanced ages remains to be determined.

Young M-dwarfs can exhibit strong levels of UV and X-ray radiation, either as a result of chromospheric and coronal activity or active accretion. This high energy radiation can drive disk dissipation via photoevaporation (e.g., \citealt{Gorti:2009}).
Nevertheless, the limited studies that have been performed to explore the disk lifetimes of substellar objects  have found disk dissipation timescales are at least as long as for solar-mass stars (see, e.g., \citealt{Luhman:2012}, \citealt{Ercolano:2011}, \citealt{Williams:2011}, and references therein).
Investigations aimed at establishing the masses of gas and dust around similarly young M stars are key to advance our understanding of disk evolution timescales and processes.

%SEDs
\begin{figure*}
\begin{center}
\includegraphics[width=8cm,angle=0]{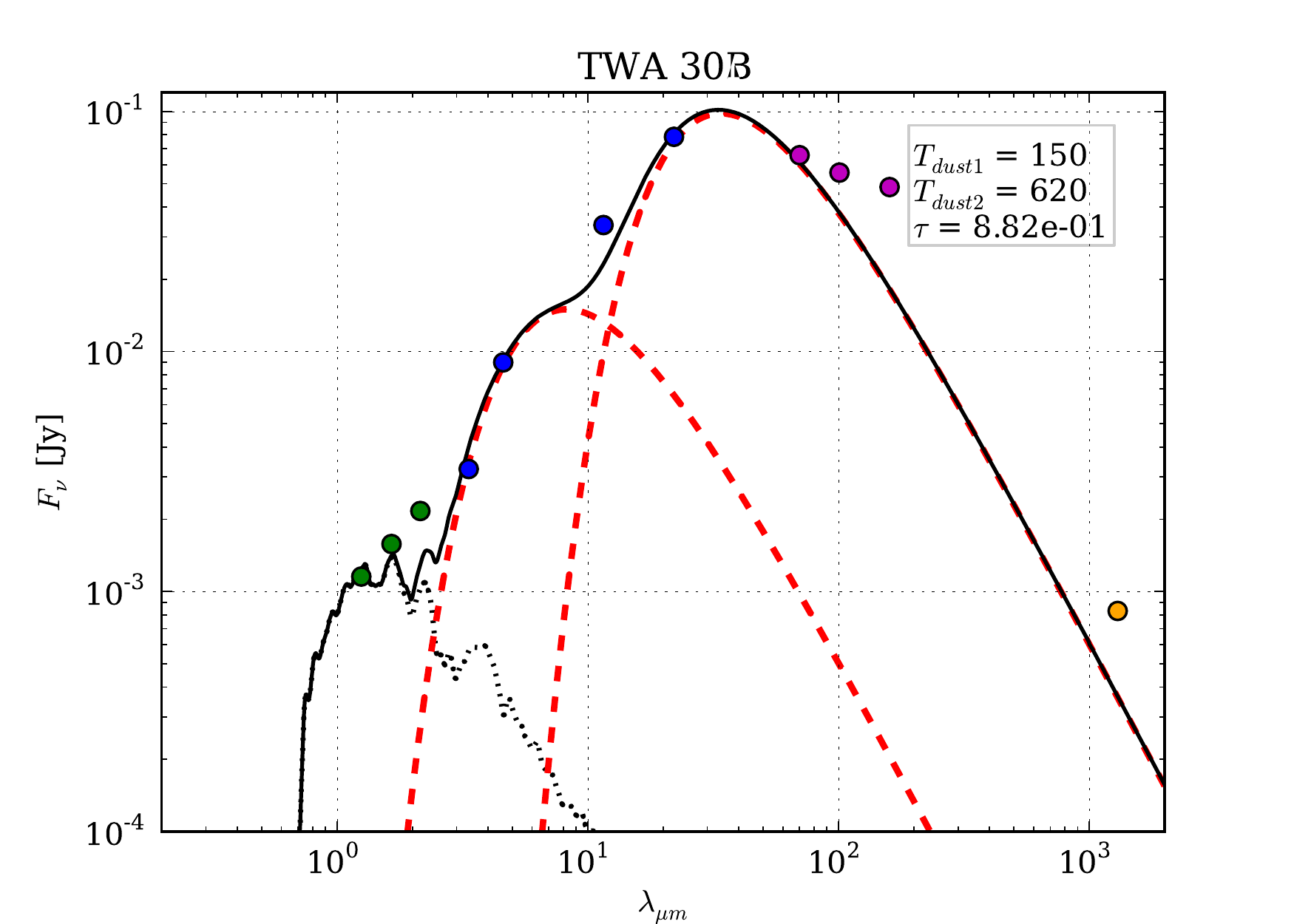}
\includegraphics[width=8cm,angle=0]{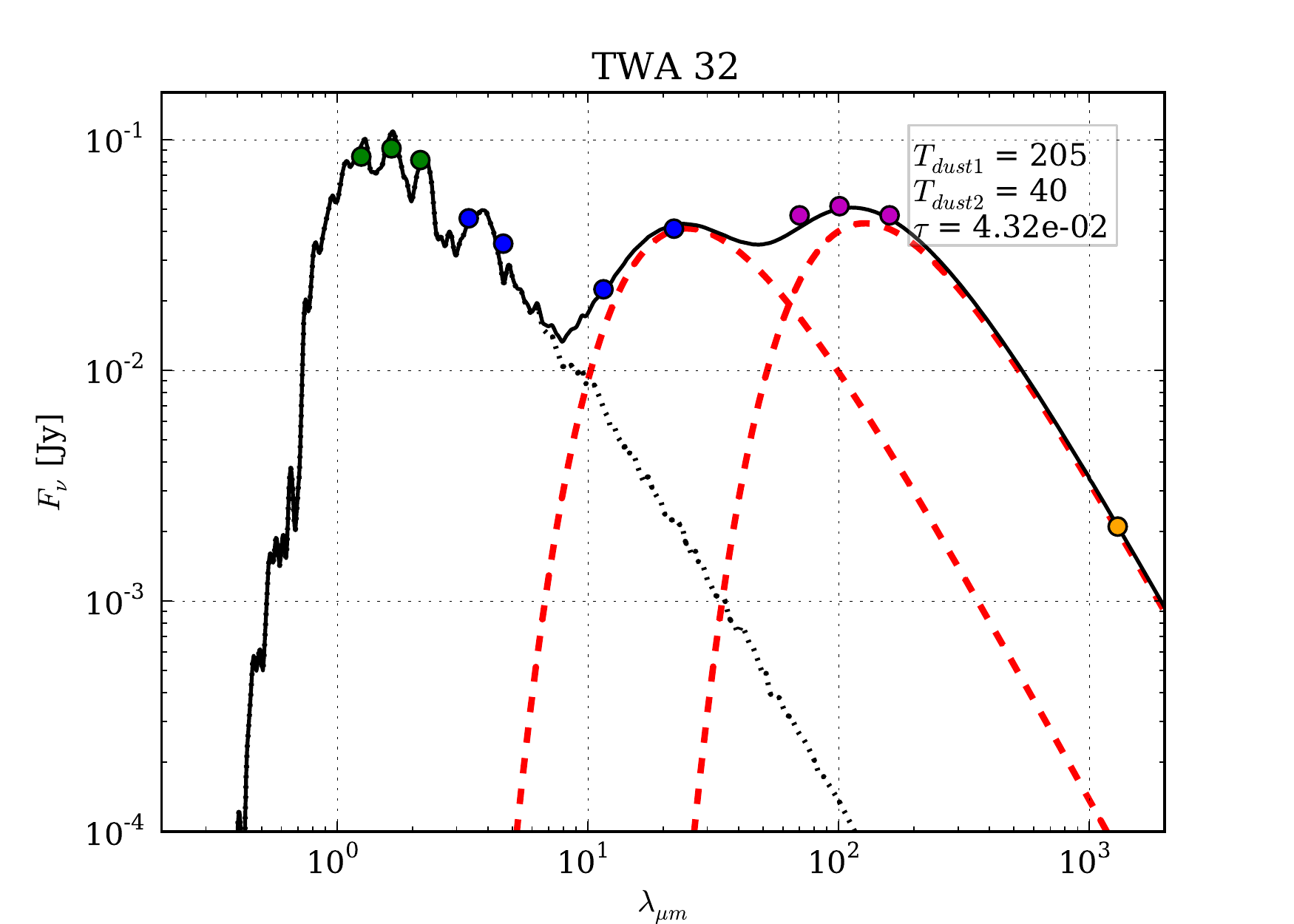}
\includegraphics[width=8cm,angle=0]{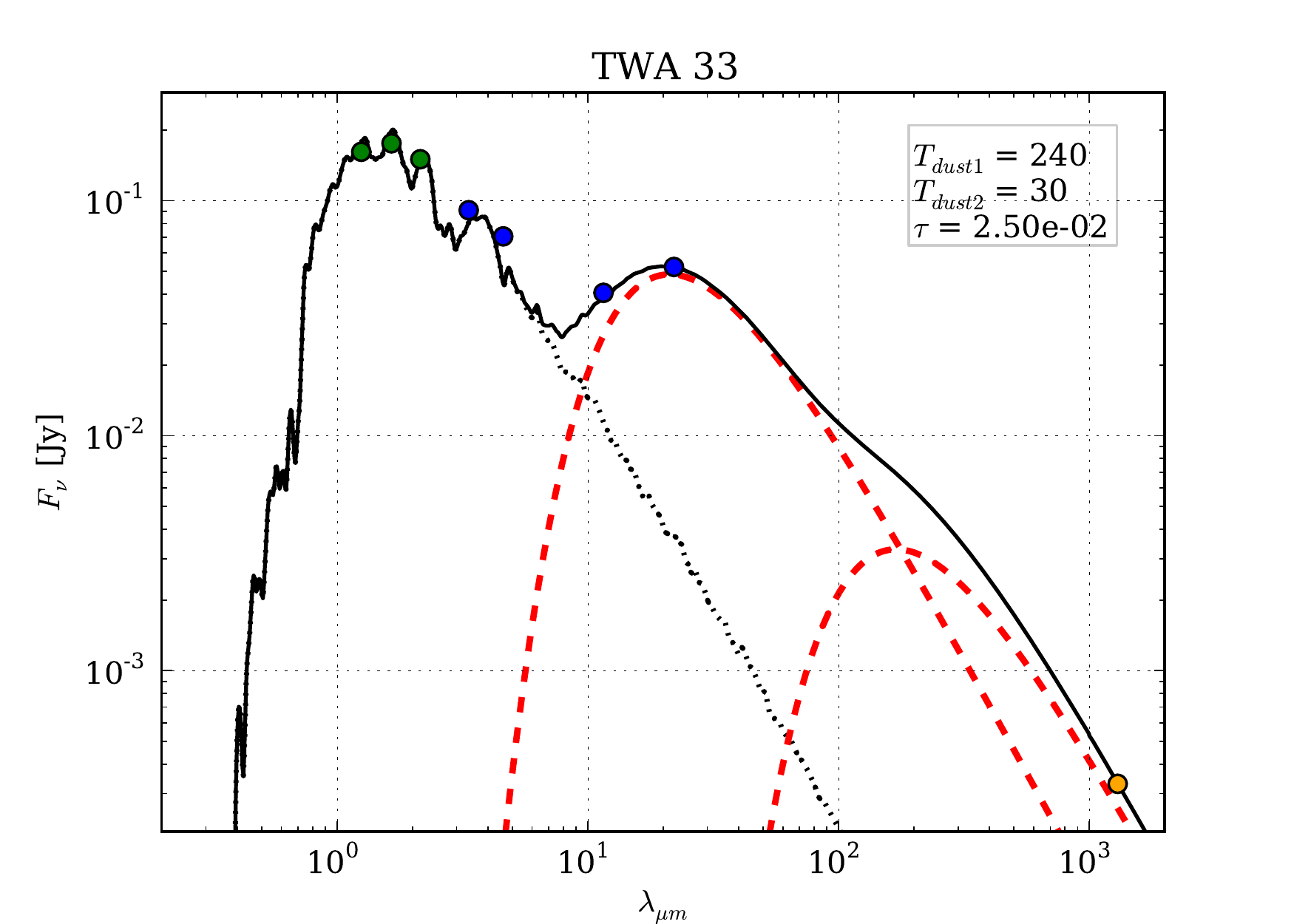}
\includegraphics[width=8cm,angle=0]{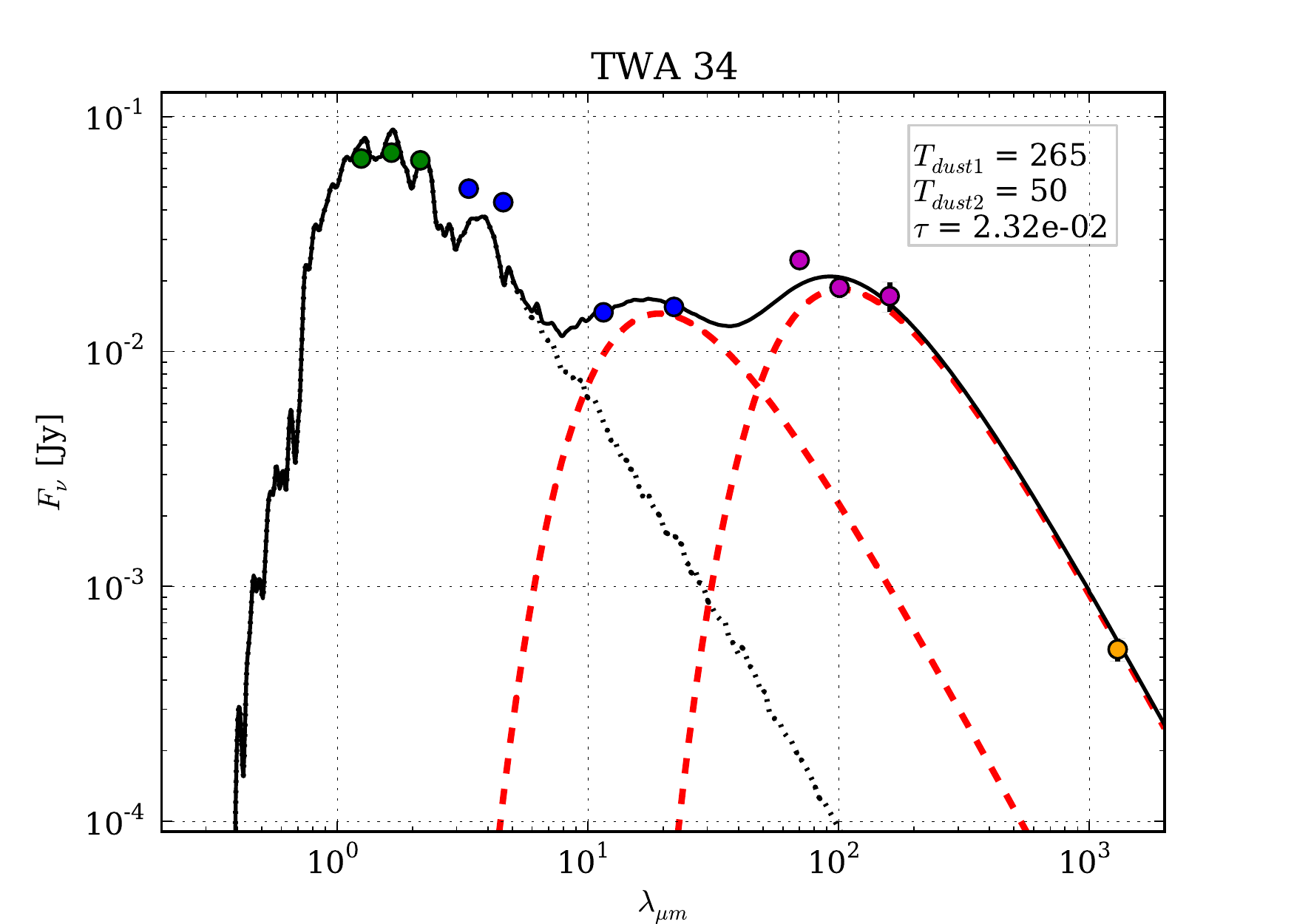}
\end{center}
\caption{Spectral energy distributions for TWA~30B, 32, 33, and 34. Data points are 2MASS (green), ALLWISE (blue), Herschel (purple; see \citealt{Liu:2015}), and ALMA (orange). Two blackbody dust disks are fit in each case. The fractional luminosities, $\tau=L_{IR}/L_{bol}$, are indicated. The poor fit to TWA~30B suggests a more complex model is required.}
\label{fig:seds}
\end{figure*}

Among the nearby young moving groups, the TWA is particularly interesting, as it represents an important evolutionary stage in the lives of protoplanetary disks. 
With an age of $\sim$8~Myr \citep{Torres:2008,Ducourant:2014}, the TWA represents an epoch coinciding with the time required for giant planet formation via core accretion (eg, \citealt{Chabrier:2014}). 
The disks in the TWA range from evidently gas-poor debris disks to at least one example of a long-lived, gas-rich, apparently primordial disk --- i.e., the disk orbiting TW~Hya itself (e.g., \citealt{Riviere:2013, Schneider:2012a, Andrews:2010}; and references therein).
Recent work has yielded many new M-dwarf members and candidate members of the TWA \citep{Looper:2007,Looper:2010a,Looper:2010b,Looper:2011,Shkolnik:2011,Rodriguez:2011,Schneider:2012b,Gagne:2014b}. 
With masses lower than $\sim$0.2~$M_\odot$, these new mid/late M-dwarfs constitute a sample of objects that allows us to probe disks among the poorly explored substellar population, and to do so at the relatively advanced age of the TWA.

\section{Observations}

We carried out an ALMA survey of 15 members or proposed members of the TWA as drawn from \citet{Looper:2007,Looper:2010a,Looper:2010b,Looper:2011,Shkolnik:2011,Rodriguez:2011,Schneider:2012b,Gagne:2014b}. 
These stars, listed in Table~\ref{tab:targets}, were chosen as targets as they constitute some of the lowest mass members suggested to date for the TWA. 
None of these had yet been observed with ALMA, though most are known to host dusty circumstellar disks as inferred via WISE and Herschel infrared excesses (eg, \citealt{Schneider:2012a,Schneider:2012b, Liu:2015}).
For those not in published IR surveys, we noted the WISE color $W1-W4$ and marked those with colors redder than 2 as having an IR excess.
Some targets, notably J1326--5022 and TWA~29, have low TWA membership probabilities ($<$10\%) according to BANYAN~II \citep{Gagne:2014a}, but were nevertheless included in our survey. 
BANYAN~II returns a 98\% likelihood of membership for TWA~31; however, other studies do not consider it a likely member (e.g., \citealt{Ducourant:2014}).

Our ALMA Cycle~2 program (2013.1.00457.S) consisted of observations of continuum dust emission at 230~GHz and observations of the $^{12}$CO(2--1) and $^{13}$CO(2--1) emission lines with a resolution of 488~kHz, corresponding to a velocity resolution of 0.6~km/s. 
We reached a sensitivity of 0.05~mJy/beam in the continuum and 5~mJy/beam per 0.6~km/s channel in $^{12}$CO and $^{13}$CO. Calibration and cleaning was performed by the ALMA staff with CASA version 4.2.2. Briggs weighting was used with robust=0.5. The final restored beam was, on average, $1.5\times0.8$\arcsec\ and corresponds to a scale of 40--80~AU at the mean distance of the stars in the TWA.

\section{Results} \label{results}

\begin{table*}
\begin{center}
\begin{tabular}{lccccccccrr}
\hline
Name & RA & Dec.\ & Sp.\ & W1-W3 & W1-W4 & IR & Distance & Flux & $M_{dust}$ & Ref. \\
 	  & & & Type         & (mag) & (mag) & Excess & (pc) & (mJy) & ($10^{-2} M_{E}$) &  \\
\hline
\hline
TWA 30B		& 11:32:18 & -30:18:31 & M4	 	& 4.96 & 7.23 & Y & $45.8\pm4.8$ & $0.83\pm0.07$ & 3.7 & 1 \\
TWA 30A		& 11:32:18 & -30:19:51 & M5	 	& 1.72 & 3.64 & Y & $45.8\pm4.8$ & $<$0.20 & $<$0.9 & 2 \\
TWA 31		& 12:07:10 & -32:30:53 & M5		& 1.80 & 3.92 & Y & $55.0\pm6.4$ & $<$0.20 & $<$1.3 & 3 \\
TWA 33		& 11:39:33 & -30:40:00 & M5 	& 1.63 & 3.25 & Y & $46.6\pm5.2$ & $0.33\pm0.03$ & 1.5 & 4  \\
TWA 34	 	& 10:28:45 & -28:30:37 & M5		& 1.35 & 2.75 & Y & $47.0\pm5.6$ & $0.54\pm0.06$ & 2.5 & 4 \\
TWA 32		& 12:26:51 & -33:16:12 & M6		& 1.77 & 3.77 & Y & $59.4\pm6.2$ & $2.10\pm0.05$ & 15.8 & 3, 5 \\
J1045-2819	& 10:45:52 & -28:19:30 & M6		 & 0.32 & $<$2.73 & ? & $52.6\pm6.4$ & $<$0.15 & $<$0.9 &  6 \\
J1111-2655	& 11:11:28 & -26:55:02 & M6		 & 0.47 & 0.93 & N & $43.8\pm5.0$ & $<$0.16 & $<$0.7 &  6 \\
J1203-3821	& 12:03:59 & -38:21:40 & M8		 & 0.80 & $<$3.09 & ? & $65.8\pm8.6$ & $<$0.15 & $<$1.4 &   6 \\
J1252-4948	& 12:52:09 & -49:48:28 & M8/9	 & 0.86 & 3.00 & Y & $91.8\pm9.4$ & $<$0.13 & $<$2.3 &  6 \\
J1106-3715	& 11:06:44 & -37:15:11 & M9		 & $<$0.88 & $<$4.02 & ? & $53.4\pm8.0$ & $<$0.12 & $<$0.7 &  6 \\
J1326-5022	& 13:26:53 & -50:22:27 & M9		 & 1.44 & $<$3.32 & ? & $71.0\pm9.0$ & $<$0.13 & $<$1.4 & 6 \\
J1247-3816	& 12:47:44 & -38:16:46 & M9		 & 2.16 & 4.27 & Y & $63.8\pm6.4$ & $<$0.15 & $<$1.3 & 7 \\
TWA 29		& 12:45:14 & -44:29:07 & M9/9.5		 & $<$0.43 & $<$3.90 & ? & $33.3\pm7.2$ & $<$0.14 & $<$0.3 & 8 \\
J1207-3900	& 12:07:48 & -39:00:04 & L0/1		 & 0.43 & $<$4.44 & ? & $60.2\pm5.2$ & $<$0.14 & $<$1.1 & 7 \\
\hline\end{tabular}
\caption{Targets for our ALMA observations.  The third column indicates if the source had prior indications of IR excess from WISE ($W1-W4>2$); "?" denotes upper limits at WISE bands $W3$, $W4$, or both.
Distances listed here have been calculated via the BANYAN II kinematic tool \citep{Gagne:2014a}.
Also listed are the continuum detections and 3-$\sigma$ upper limits. Dust masses are estimated as described in Section~\ref{results} and using $T_{dust}\sim40$~K. All sources are unresolved with ALMA. 
\newline References: (1) \citealt{Looper:2010b}, (2) \citealt{Looper:2010a}, (3) \citealt{Shkolnik:2011}, (4) \citealt{Schneider:2012b}, (5) \citealt{Rodriguez:2011}, (6) \citealt{Looper:2011}, (7) \citealt{Gagne:2014b}, (8) \citealt{Looper:2007}.
} \label{tab:targets}
\end{center}
\end{table*}

Of the 15 targets observed, four systems were detected in continuum emission. All four sources were unresolved. 
These first-time detections are listed in Table~\ref{tab:targets} along with the measured flux from the primary beam corrected images. 
Fluxes were measured by fitting a Gaussian to the cleaned image.
We also list 3-$\sigma$ limits, which are three times the RMS error estimated in each case.
The measured fluxes are consistent with emission from T$\sim$40~K dust grains as determined by modeling the spectral energy distribution (SED). 
Figure~\ref{fig:seds} shows SEDs of these 4 systems along with blackbody fits to their IR/submm excesses (see, e.g., \citealt{Schneider:2012a}). 
In the case of TWA 30B, which displays strong variability and may have complex disk structure as a result of the edge-on disk geometry (\citealt{Looper:2010b}; Principe et al., submitted), a more sophisticated model is clearly required to accurately describe its SED.

These SED models are overly simplistic; unless the emission is coming from narrow rings, the real disks will have a range of temperatures. Nevertheless, the models in Fig.~\ref{fig:seds} are useful to demonstrate the presence of cold dust in the system. The fractional luminosity, $\tau = L_{IR}/L_{bol}$, ranges from 2 to 4\% for TWA~32, 33, and 34. If we assume the continuum emission is coming from large (mm-sized) grains which radiate as blackbodies at $\sim$40~K, then the dust grains need to be located a few AU from these low-mass stars. 

In addition to the four Table~\ref{tab:targets} stars that are coincident with submm continuum sources, we detected continuum sources that are well offset (typically by $\sim$10\arcsec) from three systems (TWA 31, J1247-3816, and J1207-3900). 
Given the typical positional accuracy of ALMA observations is better than 0.1\arcsec, we conclude these are background sources, likely of extragalactic origin.

Assuming optically thin dust, we can estimate the dust mass from:
\begin{gather*}
M_{dust} = \frac{F_\nu D^2}{\kappa_\nu B_\nu(T_{dust})},
\end{gather*}
where we adopt $\kappa_\nu = 1.15$~cm$^2$/g (see \citealt{Rodriguez:2010} and references therein) and $T_{dust}=40$~K. 
We tabulate the resulting estimates of $M_d$, as well as 3-$\sigma$ upper limits, in Table~\ref{tab:targets}. 
The inferred dust masses range from $\sim$1 to 16\% of an Earth mass. 

An alternative approach to determine dust masses is to use the temperature-luminosity relationship for protoplanetary disks derived in \citet{Andrews:2013}, namely $T_d \approx 25 (L_*/L_\odot)^{1/4}$. 
While consistent with results for disks orbiting young, earlier type stars, the \citet{Andrews:2013} relationship does not necessarily hold for very low-mass objects such as those in Table~\ref{tab:targets} (van der Plas et al., in prep). 
Young M5 stars have log~$L_*/L_\odot \approx -2$, which suggests a dust temperature of $\sim$8~K. 
This is 5 times lower than the temperatures assumed in the SED model and, under the same assumptions, implies dust masses that are 5 times higher. 

In addition to obtaining flux upper limits for the individual continuum non-detections, we created a stacked image by averaging non-detections, to assess whether a significant fraction of them might display emission at levels just below detectability. No detection is apparent in this average image; we obtain a 3-$\sigma$ upper limit of 0.05 mJy/beam.

Among our sample only TWA~34 displays detectable $^{12}$CO emission, which we discuss in Section~\ref{twa34co}.
For those systems in our sample with no CO detections, we can infer a 3-$\sigma$ upper limit of $\sim$0.002~$M_E$ in molecular gas following the prescription in Section~\ref{twa34co} and assuming optically thin $^{12}$CO, CO:H$_2$ of 10$^{-4}$, and a distance of $\sim$60~pc.

\subsection{TWA~34: CO Detection}\label{twa34co}

\begin{figure}
\begin{center}
\includegraphics[width=8cm,angle=0]{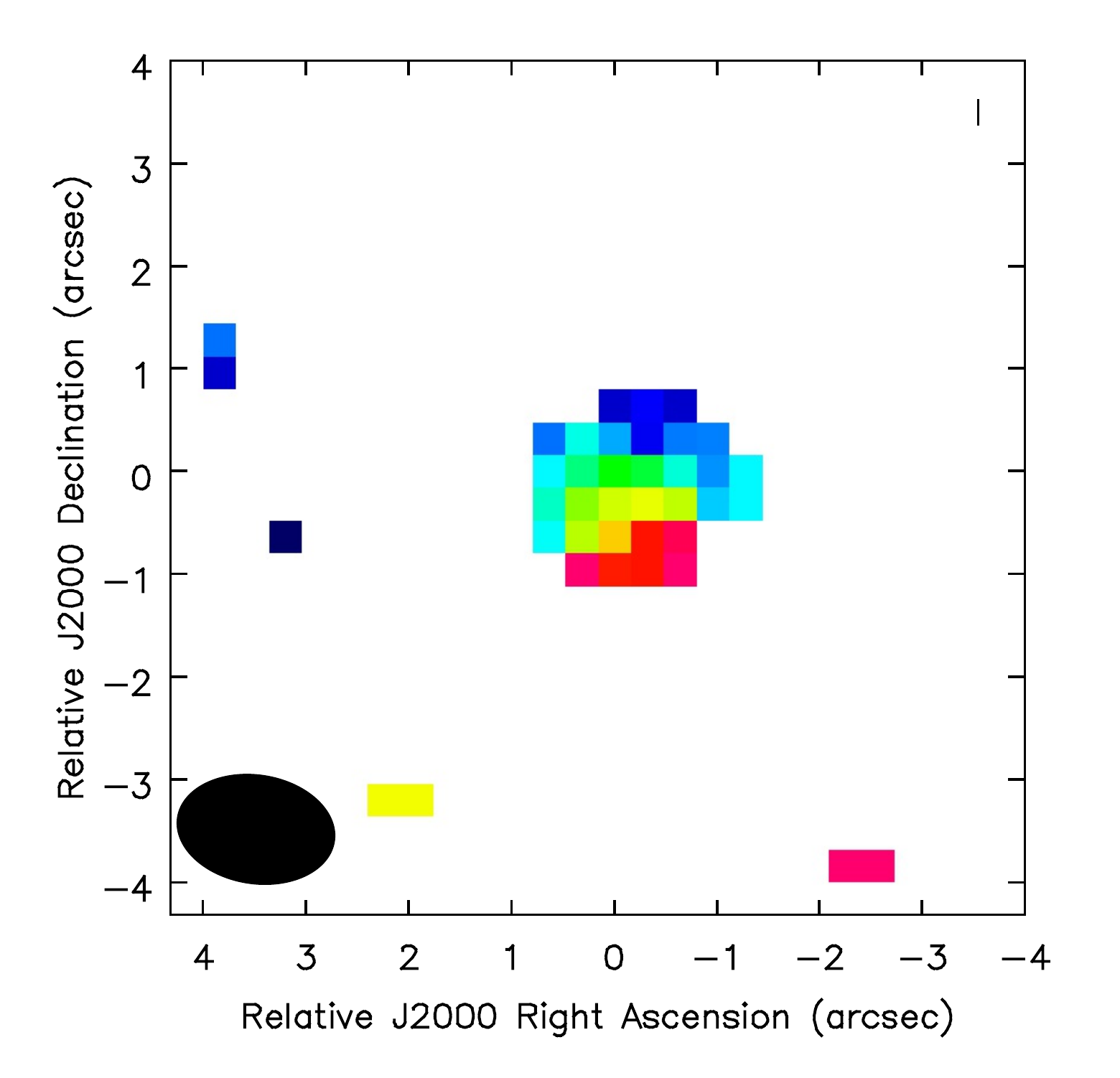}
\end{center}
\caption{Velocity map for the $^{12}$CO(2--1) emission in TWA~34. }
\label{fig:mom1}
\end{figure}

Among the Table~\ref{tab:targets} systems, only TWA~34 shows evidence of $^{12}$CO emission. 
Although the $^{12}$CO emission is unresolved for each individual velocity channel, we find the centroid changes with velocity, allowing us to generate the first moment map in Figure~\ref{fig:mom1}.
Figure~\ref{fig:mom1} suggests that TWA~34 is orbited by a molecule-rich disk viewed at intermediate to high inclination, with North-South rotation.

%moment map

\begin{figure}
\begin{center}
\includegraphics[width=8cm,angle=0]{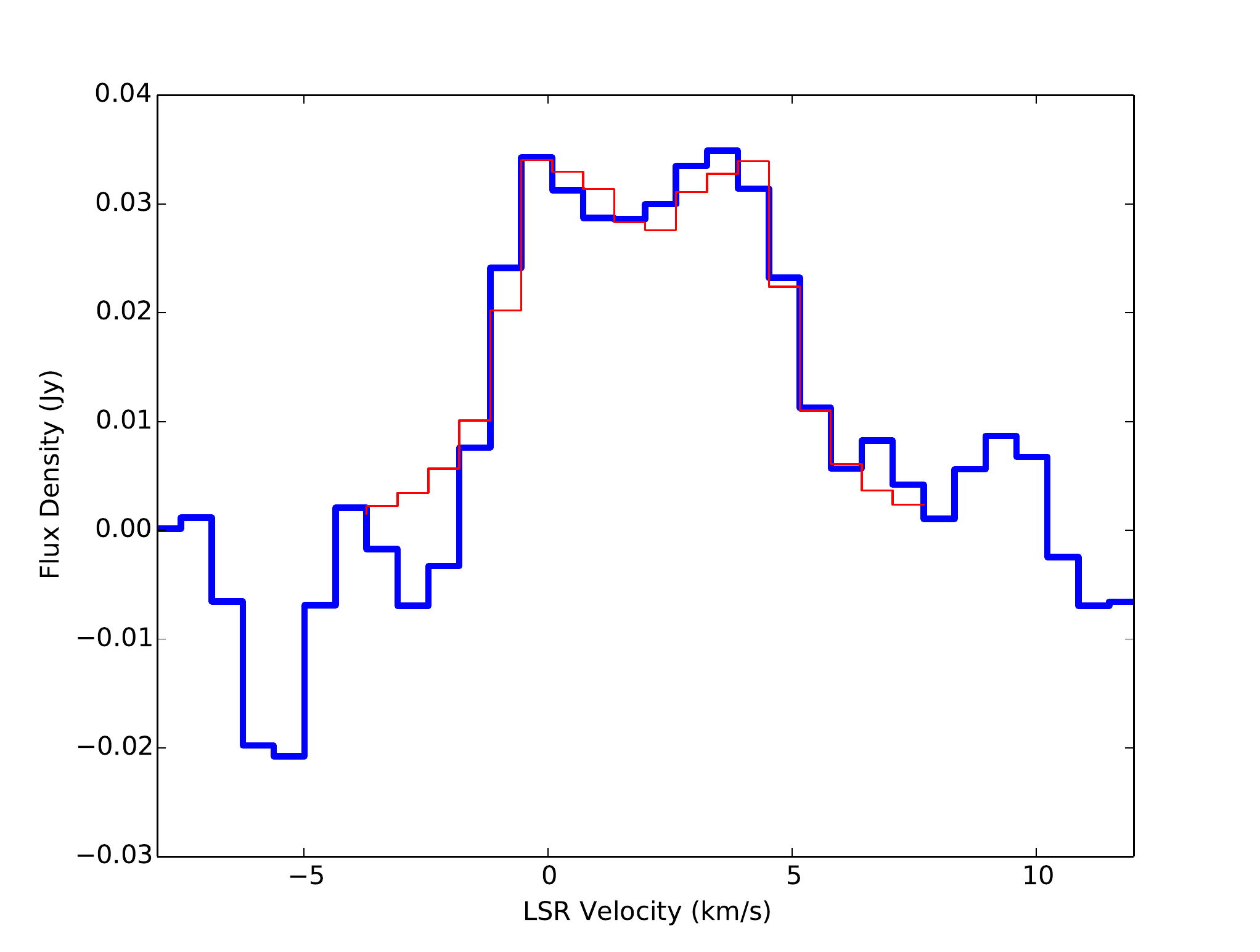}
\end{center}
\caption{$^{12}$CO(2--1) emission line profile of TWA~34, the only target in our sample with detected CO gas. The red, thin line represents the best-fit Keplerian profile.}
\label{fig:spec}
\end{figure}

We show in Figure~\ref{fig:spec} the integrated line profile of the CO emission, Hanning smoothed with a kernel size of 3 channels. 
This double-peaked CO line profile is indicative of Keplerian rotation. Hence, to characterize this emission, we fit a parametrized Keplerian model as described in \citet{Kastner:2008a}. That is, we fit a parametric line profile function described by: 
\begin{align*}
F = \left\{
  \begin{array}{lr}
    F_0 \ ((v-v_0)/v_d)^{3q-5} &  |v-v_0|>v_d \\
    F_0 \ ((v-v_0)/v_d)^{p_d} &  |v-v_0|<v_d
  \end{array}
\right.
\end{align*}
where $F_0$ is the peak line intensity, $\nu_0$ is the rest frequency in the star/disk system frame, $v_d$ is the projected rotational velocity near the outer edge of the disk, and $p_d$ and $q$ are quasi-physical disk parameters (see \citealt{Kastner:2008a} for details).
We fix $q$=0.5 and $p_d$=0.1 for simplicity.  The resulting best-fit parametric line profile is displayed in Figure~\ref{fig:spec}. 
We obtain a peak intensity of $0.034\pm0.002$~Jy, a systemic velocity of $2.3\pm0.1$~km/s in the LSR frame, $v_d$ of $2.49\pm0.09$km/s, and an integrated intensity of $0.34\pm0.03$~Jy~km/s. The parameter $v_d$ can be used to estimate the outer radius of the disk as detected in CO emission, $R_d$, from
\begin{gather*} 
v_d^2 = G M_* / R_d
\end{gather*}
where $M_*$ is the mass of the star, 0.08~$M_\odot$ \citep{Baraffe:2015}.
We thereby estimate the CO disk orbiting TWA~34 is $\sim$11 AU in radius.
This suggests that the disk around TWA~34 is more compact than those seen around younger brown dwarfs and low-mass stars in Taurus \citep{Ricci:2014}.
We note that the CO emission in Figure~\ref{fig:mom1} is marginally resolved and suggests a larger CO radius of $\sim$20--40~AU. This discrepancy in the CO outer radius estimates can also be seen in other disks when comparing single-dish line measurements to resolved interferometric imaging (e.g., \citealt{Kastner:2008b, Rodriguez:2010, Sacco:2014, Huelamo:2015}). 
Higher resolution CO imaging will allow an independent measurement of the disk size, which can in turn be used to accurately estimate the mass of TWA~34.

We estimate a gas mass of  $\sim$0.2~$M_E$ following the prescriptions of \citet{Zuckerman:2008}, \citet{Kastner:2008a}, \citet{Rodriguez:2010}, and references therein. We assume optically thin $^{13}$CO, a $^{12}$C:$^{13}$C ratio of 89, a CO:H$_2$ ratio of 10$^{-4}$, temperature of $\sim$40~K, and a 3-$\sigma$ $^{13}$CO upper limit of $\sim$0.03~Jy~km/s.  The $^{13}$CO upper limit is determined assuming a linewidth identical to that of $^{12}$CO.
This estimate for the gas mass is $\sim$7 times higher than that for the dust mass.
This ratio is comparable to other evolved molecular gas disks such as TW~Hya and V4046~Sgr, as well as some disks in the much younger ($\sim$1--3~Myr) Taurus star-forming region (e.g., \citealt{Sacco:2014,Williams:2014}). 
As these and many other previous studies suggest, such low inferred gas-to-dust ratios could be indicative of gas removal through accretion or photoevaporation, or could merely reflect overestimates of the CO:H$_2$ ratio; due, for example, to freezing out of CO on cold dust grains. These various processes result in large uncertainties when estimating the gas mass of disks based on CO measurements.

% Spectra

\subsection{TWA 34: System Velocity}

From our best-fit Keplerian profile (see prior section), we have obtained an estimate for the systemic velocity of TWA~34.
In the barycentric frame of reference, this velocity corresponds to $13.3\pm0.1$~km/s. 
This is the first accurate radial velocity measurements available for TWA~34 (see also \citealt{Murphy:2015}).
At a distance of $47.0\pm5.6$~pc and using the proper motions listed in PPMXL \citep{Roeser:2010}, we can estimate UVW velocities of $-11.0\pm1.6$, $-16.2\pm0.6$, $-3.9\pm1.4$~km/s. 
These agree very well with the average velocity of the TWA ($-9.87$, $-18.06$, $-4.52$~km/s; \citealt{Malo:2013}).
This new velocity measurement thereby further supports the conclusion that TWA~34 is a member of the TWA.

\section{Discussion}

Among the 15 low-mass TWA members and candidate members listed in Table~\ref{tab:targets}, only 4 yielded ALMA continuum detection at 1.3 mm, despite the fact that many show some evidence of warm circumstellar dust.
In the absence of cold dust grains, the warm dust grains detected by WISE would have 1.3~mm emission $<$0.1~mJy, which is below the sensitivity of our ALMA observations.
Because we are only sensitive to cold mm-sized grains, our observations appear to demonstrate that the presence of warm circumstellar dust does not necessarily imply cold dust is also present in the system.
This is in agreement with prior studies of M stars (see, e.g., \citealt{Lestrade:2009} and references therein) that have found the incidence of cold disks to be much lower around such stars than around higher mass stars.
Although the ALMA non-detections appear to indicate no cold grains exist, an alternative explanation is that any surviving grains in the outer disk have already grown to cm size or larger and become invisible at 1.3~mm wavelengths \citep{Ricci:2010a,Ricci:2010b,Mohanty:2013}.

The continuum non-detections suggest dust masses of about $10^{-2} M_E$ or less. 
This is similar to what has been observed for other $\sim$10~Myr-old debris disks around M stars (see \citealt{Wyatt:2008} and references therein).
We were more sensitive to molecular gas masses and achieved a limit of a few times $10^{-3} M_E$ for gas in H$_2$, assuming CO/H$_2$ of $10^{-4}$. As the case for TWA~34 and other disks shows, the gas-to-dust ratio is unlikely to be $\sim$100 as in the ISM (see \citealt{Williams:2014}). 
It appears likely that by the age of the TWA, the gas in a typical M star's disk has in general been efficiently removed, even in cases where a significant mass of primordial dust has survived.
However, studies have identified signatures of on-going gas accretion in some of these systems, for example around TWA~30A, 30B, and 31 \citep{Looper:2010a,Looper:2010b, Shkolnik:2011}.
Hence, at least in certain cases, it is likely that disk CO gas has frozen out onto dust grains, suppressing the gas-phase CO abundance.

\section{Conclusions}

We have carried out an ALMA survey of 15 low-mass TWA members and candidates to search for molecular gas in the form of $^{12}$CO and $^{13}$CO as well as provide constraints on continuum dust emission. 
Among systems targeted, four (TWA~30B, 32, 33, and 34) have detected dust emission consistent with the existence of cold dust grains in the disk. 
Circumstellar dust grain temperatures of $\sim$40 K are consistent with the mid-infrared to submm SEDs for these systems.
All continuum sources are unresolved. 
While most of our sample shows indications of warm dust based on WISE measurements, the ALMA non-detections suggest any cold grains present in the outer disk may have already grown to cm size or larger.

Only one system, TWA~34, shows signatures of molecular gas in its disk in the form of $^{12}$CO (2--1) emission. 
The $^{12}$CO emission has velocity structure indicative of Keplerian rotation. The systemic velocity for the system, as determined from the CO detection, is consistent with membership in the TWA.
Among the sample of known $\sim$7--10 Myr-old star/disk systems, TWA~34, at just $\sim$50 pc from Earth, is the lowest mass star thus far identified as harboring cold molecular gas in an orbiting disk.

\begin{acknowledgements}
This paper makes use of the following ALMA data: ADS/JAO.ALMA\#2013.1.00457.S. ALMA is a partnership of ESO (representing its member states), NSF (USA) and NINS (Japan), together with NRC (Canada), NSC and ASIAA (Taiwan), and KASI (Republic of Korea), in cooperation with the Republic of Chile. The Joint ALMA Observatory is operated by ESO, AUI/NRAO and NAOJ.
We thank our referee, Greg Herczeg, for the detailed and useful review of our manuscript.
D.R.R. acknowledges support from FONDECYT grant 3130520.
G.v.d.P. acknowledges support from FONDECYT grant 3140393.
D.P. acknowledges support from FONDECYT grant 3150550.
G.v.d.P. acknowledges support from FONDECYT grant 3140393 and by the Millennium Nucleus RC130007 (Chilean Ministry of Economy).
J.K.'s research on young stars near Earth is supported by National Science Foundation grant AST-1108950 and NASA Astrophysics Data Analysis Program grant NNX12H37G, both to RIT.
S.M. acknowledges the support of the STFC grant ST/K001051/1
\end{acknowledgements}


\begin{thebibliography}{}

\bibitem[Andrews et al.(2010)]{Andrews:2010} Andrews, S.~M., Czekala, I., Wilner, D.~J., et al.\ 2010, \apj, 710, 462 % truncated disks of HD98800 and Hen 3-600
\bibitem[Andrews et al.(2013)]{Andrews:2013} Andrews, S.~M., Rosenfeld, K.~A., Kraus, A.~L., \& Wilner, D.~J.\ 2013, \apj, 771, 129 

\bibitem[Baraffe et al.(2015)]{Baraffe:2015} Baraffe, I., Homeier, D., Allard, F., \& Chabrier, G.\ 2015, \aap, 577, A42 % pre-MS models

\bibitem[Chabrier et al.(2014)]{Chabrier:2014} Chabrier, G., Johansen, A., Janson, M., \& Rafikov, R.\ 2014, Protostars and Planets VI, 619

\bibitem[Ducourant et al.(2014)]{Ducourant:2014} Ducourant, C., Teixeira, R., Galli, P.~A.~B., et al.\ 2014, \aap, 563, A121 %TW Hya age

\bibitem[Ercolano et al.(2011)]{Ercolano:2011} Ercolano, B., Bastian, N., Spezzi, L., \& Owen, J.\ 2011, \mnras, 416, 439 % disk lifetimes around late-type stars

\bibitem[Gagn{\'e} et al.(2014a)]{Gagne:2014a} Gagn{\'e}, J., Lafreni{\`e}re, D., Doyon, R., Malo, L., \& Artigau, {\'E}.\ 2014, \apj, 783, 121 
\bibitem[Gagn{\'e} et al.(2014b)]{Gagne:2014b} Gagn{\'e}, J., Faherty, J.~K., Cruz, K., et al.\ 2014, \apjl, 785, LL14 

\bibitem[Gorti \& Hollenbach(2009)]{Gorti:2009} Gorti, U., \& Hollenbach, D.\ 2009, \apj, 690, 1539 % UV/Xray photoevaporation of disks

\bibitem[Hu{\'e}lamo et al.(2015)]{Huelamo:2015} Hu{\'e}lamo, N., de Gregorio-Monsalvo, I., Macias, E., et al.\ 2015, \aap, 575, L5 %T Cha with ALMA

\bibitem[Kastner et al.(1997)]{Kastner:1997} Kastner, J.~H., Zuckerman, B., Weintraub, D.~A., \& Forveille, T.\ 1997, Science, 277, 67
\bibitem[Kastner et al.(2008)]{Kastner:2008a} Kastner, J.~H., Zuckerman, B., \& Forveille, T.\ 2008, \aap, 486, 239 % BP Psc
\bibitem[Kastner et al.(2008)]{Kastner:2008b} Kastner, J.~H., Zuckerman, B., Hily-Blant, P., \& Forveille, T.\ 2008, \aap, 492, 469 % V4046Sgr
\bibitem[Kastner et al.(2010)]{Kastner:2010} Kastner, J.~H., Hily-Blant, P., Sacco, G.~G., Forveille, T., \& Zuckerman, B.\ 2010, \apjl, 723, L248 %MP Mus

\bibitem[Lagrange et al.(2010)]{Lagrange:2010} Lagrange, A.-M., Bonnefoy, M., Chauvin, G., et al.\ 2010, Science, 329, 57 % beta Pic planet
\bibitem[Lestrade et al.(2009)]{Lestrade:2009} Lestrade, J.-F., Wyatt, M.~C., Bertoldi, F., Menten, K.~M., \& Labaigt, G.\ 2009, \aap, 506, 1455
\bibitem[Liu et al.(2015)]{Liu:2015} Liu, Y., Herczeg, G.~J., Gong, M., et al.\ 2015, \aap, 573, A63

\bibitem[Looper et al.(2007)]{Looper:2007} Looper, D.~L., Burgasser, A.~J., Kirkpatrick, J.~D., \& Swift, B.~J.\ 2007, \apjl, 669, L97 
\bibitem[Looper et al.(2010a)]{Looper:2010a} Looper, D.~L., Mohanty, S., Bochanski, J.~J., et al.\ 2010, \apj, 714, 45
\bibitem[Looper et al.(2010b)]{Looper:2010b} Looper, D.~L., Bochanski, J.~J., Burgasser, A.~J., et al.\ 2010b, \aj, 140, 1486 
\bibitem[Looper(2011)]{Looper:2011} Looper, D.~L.\ 2011, Ph.D.~Thesis, University of Hawai'i 

\bibitem[Luhman \& Mamajek(2012)]{Luhman:2012} Luhman, K.~L., \& Mamajek, E.~E.\ 2012, \apj, 758, 31 %disk population & timescales in Upper Sco

\bibitem[Malo et al.(2013)]{Malo:2013} Malo, L., Doyon, R., Lafreni{\`e}re, D., et al.\ 2013, \apj, 762, 88 

%\bibitem[Mamajek et al.(2013)]{Mamajek:2013} Mamajek, E.~E., Bartlett, J.~L., Seifahrt, A., et al.\ 2013, \aj, 146, 154

\bibitem[Marois et al.(2008)]{Marois:2008} Marois, C., Macintosh, B., Barman, T., et al.\ 2008, Science, 322, 1348 %HR8799 planets

\bibitem[Mohanty et al.(2013)]{Mohanty:2013} Mohanty, S., Greaves, J., Mortlock, D., et al.\ 2013, \apj, 773, 168

\bibitem[Murphy et al.(2015)]{Murphy:2015} Murphy, S.~J., Lawson, W.~A., \& Bento, J.\ 2015, MNRAS, in press (arXiv:1507.08002)
%\bibitem[Pecaut \& Mamajek(2013)]{Pecaut:2013} Pecaut, M.~J., \& Mamajek, E.~E.\ 2013, \apjs, 208, 9 

%\bibitem[Riedel et al.(2014)]{Riedel:2014} Riedel, A.~R., Finch, C.~T., Henry, T.~J., et al.\ 2014, arXiv:1401.0722

\bibitem[Ricci et al.(2010a)]{Ricci:2010a} Ricci, L., Testi, L., Natta, A., et al.\ 2010, \aap, 512, A15
\bibitem[Ricci et al.(2010b)]{Ricci:2010b} Ricci, L., Testi, L., Natta, A., \& Brooks, K.~J.\ 2010, \aap, 521, A66 
\bibitem[Ricci et al.(2014)]{Ricci:2014} Ricci, L., Testi, L., Natta, A., et al.\ 2014, \apj, 791, 20 

\bibitem[Riviere-Marichalar et al.(2013)]{Riviere:2013} Riviere-Marichalar, P., Pinte, C., Barrado, D., et al.\ 2013, \aap, 555, A67 %gas and dust in TWA with Herschel

\bibitem[Rodriguez et al.(2010)]{Rodriguez:2010} Rodriguez, D.~R., Kastner, J.~H., Wilner, D., \& Qi, C.\ 2010, \apj, 720, 1684
\bibitem[Rodriguez et al.(2011)]{Rodriguez:2011} Rodriguez, D.~R., Bessell, M.~S., Zuckerman, B., \& Kastner, J.~H.\ 2011, \apj, 727, 62 

\bibitem[Roeser et al.(2010)]{Roeser:2010} Roeser, S., Demleitner, M., \& Schilbach, E.\ 2010, \aj, 139, 2440 % PPMXL

\bibitem[Sacco et al.(2014)]{Sacco:2014} Sacco, G.~G., Kastner, J.~H., Forveille, T., et al.\ 2014, \aap, 561, A42 %T Cha

\bibitem[Schneider et al.(2012a)]{Schneider:2012a} Schneider, A., Melis, C., \& Song, I.\ 2012a, \apj, 
754, 39 % TWA 30,31,32 disks
\bibitem[Schneider et al.(2012b)]{Schneider:2012b} Schneider, A., Song, I., Melis, C., Zuckerman, B., \& 
Bessell, M.\ 2012b, \apj, 757, 163 % TWA 33, 34

\bibitem[Shkolnik et al.(2011)]{Shkolnik:2011} Shkolnik, E.~L., Liu, M.~C., Reid, I.~N., Dupuy, T., \& Weinberger, A.~J.\ 2011, \apj, 727, 6 

%\bibitem[Teixeira et al.(2008)]{Teixeira:2008} Teixeira, R., Ducourant, C., Chauvin, G., et al.\ 2008, \aap, 489, 825

\bibitem[Torres et al.(2008)]{Torres:2008} Torres, C.~A.~O., Quast, G.~R., Melo, C.~H.~F., \& Sterzik, M.~F.\ 2008, Handbook of Star Forming Regions, Volume II, 757 

\bibitem[Williams \& Cieza(2011)]{Williams:2011} Williams, J.~P., \& Cieza, L.~A.\ 2011, \araa, 49, 67 % protoplanetary disks and their evolution
\bibitem[Williams \& Best(2014)]{Williams:2014} Williams, J.~P., \& Best, W.~M.~J.\ 2014, \apj, 788, 59 %gas/dust ratio for Taurus and other disks

\bibitem[Wyatt(2008)]{Wyatt:2008} Wyatt, M.~C.\ 2008, \araa, 46, 339 

\bibitem[Zuckerman \& Song(2004)]{ZS04} Zuckerman, B., \& Song, I.\ 2004, \araa, 42, 685
\bibitem[Zuckerman et al.(2008)]{Zuckerman:2008} Zuckerman, B., Melis, C., Song, I., et al.\ 2008, \apj, 683, 1085 %BP Psc

\end{thebibliography}
\end{document}